\documentclass[a4paper,aps,pra,showpacs,preprintnumbers]{revtex4}

\usepackage{graphics,graphicx,epsfig}
\usepackage[centertags]{amsmath}
\usepackage{amsfonts}
\usepackage{amssymb}
\usepackage[english]{babel}
\usepackage{graphicx}
\usepackage{amsthm,color}

\begin{document}



\def\BE{\begin{equation}}
\def\EE{\end{equation}}
\def\BY{\begin{eqnarray}}
\def\BEA{\begin{eqnarray}}
\def\EY{\end{eqnarray}}
\def\EEA{\end{eqnarray}}
\def\L{\label}
\def\nn{\nonumber}
\def\({\left (}
\def\){\right)}
\def\<{\langle}
\def\>{\rangle}
\def\[{\left [}
\def\]{\right]}
\def\o{\overline}
\def\BA{\begin{array}}
\def\EA{\end{array}}
\def\ds{\displaystyle}
\def\c{^\prime}
\def\cc{^{\prime\prime}}

\title{High-speed quantum memory with thermal motion of atoms}
\author{K.~Tikhonov, T.~Golubeva, Yu.~Golubev}
\affiliation{Saint Petersburg State University,\\ 198504 St.~Petersburg, Petershof, ul. Ul'yanovskaya, 1, Russia}

\date{\today}

\begin{abstract}
We discuss the influence of atomic thermal motion on the efficiency of multimode quantum memory in two configurations: over the free expand of atoms
cooled beforehand in a magneto-optical trap, and over complete mixing of atoms in a closed cell at room temperature. We consider the high-speed
quantum memory, and assume that writing and retrieval are short enough, and the displacements of atoms during these stages are negligibly small. At
the same time we take in account thermal motion during the storage time, which, as well known, must be much longer than durations of all the other
memory processes for successful application of memory cell in communication and computation. We will analyze this influence in terms of eigenmodes of
the full memory cycle and show that distortion of the eigenmodes, caused by thermal motion, leads to the efficiency reduction. We will demonstrate,
that in the multimode memory this interconnection has complicated character.
\end{abstract}

\pacs{42.50.Gy, 42.50.Ct, 32.80.Qk, 03.67.-a}%

\maketitle

\section{Introduction}

Over the last decade various protocols of quantum memory, based on the interaction between signal and driving light pulses with an ensemble of
immobilized atoms were proposed \cite{HammererReview,LvovskyREview,Simon2007,Chou2007,Chen2008}. Certainly, the approximation of motionless atoms is
natural, when one implements a quantum memory protocol on the impurities in crystals \cite{Tittel2010}. There atomic motion is restricted by the
nodes of a crystal lattice, and any spatial fluctuations are negligibly small. Besides, this approximation provides us with reliable results, if a
full memory cycle from the beginning of a writing stage and until the end of a signal retrieval (including the storage time) is short enough, and the
root-mean-square velocity is relatively small \cite{Buchler2011}.

The approximation of motionless atoms plays a significant role in analysis of multimode memory process because it allows us to follow the conversion
of a time dependance of signal into a spatial distribution of a collective spin at the writing stage. Then during the storage time this spatial
distribution keeps unchangeable, and after that, is converted back into a time profile of signal at retrieval.

However, since the purpose of quantum memory is long-term storage of information in the generated spatial coherence mode, then to adequate assessment
of the potential of such protocols we have to introduce thermal motion of atoms in a model and estimate the influence of "blur"\; on a spatial
distribution of collective coherence.

The experiments, related with light slowing, storage and manipulation on the cells with warm atomic vapors
\cite{Buchler2011,Polzik2011,Camacho2008,Novikova2011,Tabosa2014,Lvovsky2014}, are quite attractive for their comparatively simple treatment.
Notably, it is much easier to create an atomic ensemble with big number of atoms, when there is no need in its deep cooling. Moreover, a
concentration of atoms in the ensemble can be well controlled by adjusting its temperature \cite{Novikova2012}. An important characteristic of the
memory is its scalability \cite{Tittel2014,Polzik2015}. In this respect, the ensembles of room temperature atoms are more promising than cold ones,
which require extended cooling apparatus.

Thermal motion of atoms is reflected in two phenomena. First of all, it is the Doppler light shift. As a result, the medium has a Voigt absorption
profile (the convolution of Gaussian and Lorentzian profiles), what can dramatically change the character of the light-matter interaction due to its
inhomogeneous nature. However, because of collective properties of the atomic ensemble, the Lorentzian profile would be determined not by the
spontaneous decay rate of a single atom $\gamma$, but by the decay rate of the ensemble as a whole $d\gamma$, where $d$ is the optical depth. It is
the width of the profile compared with the Doppler width. That is why in the memory process the influence of thermal motion on the absorption profile
would be significantly suppressed. Furthermore, the broadening of the two-photon spin transition can be eliminated by working in collinear geometry
(e. g., co-propagating signal and driving fields) because here the transition is caused by two consequent processes - the absorption with following
the re-emission, which have equal but opposite frequency shifts.

The second phenomena induced by thermal motion, is a dependence of time of the collective spin. It occurs, because replacement of atoms leads to
reshaping of the spin distribution, which can significantly vary the memory process on each stage -- writing, retrieval, and especially storage. In
particular, thermal motion would change the spatial coherence modes at storage, that would be reflected on an optimization mechanism of the memory
efficiency \cite{Gorshkov1,Gorshkov2,Vasilyev,GrodeckaGrad,Nunn2008}.

In this article we will study the influence of atomic motion on the efficiency of multimode quantum memory. We will assume that writing and retrieval
are short enough, and therefore we can neglect any displacements of atoms on these stages. At the same time we will take in account thermal motion
during the storage time, which, as well known, must be much longer than durations of all the other memory processes for successful application of
memory cell in communication and computation. We will analyze this influence in terms of eigenmodes of the full memory cycle and show that distortion
of eigenmodes, caused by thermal motion, leads to the efficiency reduction. We will demonstrate, that in the multimode memory this interconnection
has complicated character, and prove it with the numerical calculation.

We will discuss two configurations of memory cells with atomic motion, related with the different experimental approaches. The first one assumes,
that a signal is mapped on cold atoms, prepared in a magneto-optical trap \cite{Chou2007,Chen2008,Kimble,Kuzmich,Schmiedmayer,Vuletic}. At the
beginning of writing the magneto-optical trap is switched off, and the atomic cloud begins to expand freely. The mean temperature of the ensemble is
much higher than the degeneracy temperature, so that the atomic motion satisfy Maxwell-Boltzmann statistics \cite{LandauLifshitz}. The interaction
between atoms as well as their collisions with the walls of the cell can be neglected in the assumption that the mean free path of each atom in the
cloud is much higher than its average displacement during the storage time $T_s$.  Such situation occurs, for example, in experiments
\cite{LauratQM1, LauratQM2}, where authors explore the quantum memory protocol on cesium vapors with the concentration about $10^6$ particles in
mm$^3$ and the average temperature about $100\; \mu K$.

In the second configuration we consider a room temperature atomic vapor inside the cell extended in the longitudinal direction (along the signal and
driving pulse propagation direction) and narrow in the transverse direction, so that transverse degrees of freedom are absent. We assume that during
the storage time all atoms in the cell are mixed completely, so that "which atom"\; information completely erases and the spatial spin distribution
formed on the writing stage becomes uniform. Such configuration of cells (with spin preserving coating deposited on the walls) was proposed in
\cite{Polzik2015}, where authors investigate experimentally the Raman-type quantum memory based on room temperature Cs-atoms. Note, the authors of
cited article placed the cells inside the cavity that eliminated the spatial aspect there.

The article is organized as follows. In Section \ref{II}А we discuss the physical model of high-speed quantum memory and derive the main equations
and their solutions for writing, storage and readout stages. Then, in Section \ref{II}B the eigenfunctions of full memory cycle are analyzed as well
as the corresponding eigenvalues. Here we define the response functions, which describe a spatial distribution of coherence formed at the end of the
write process, when one of the eigenfunctions is incident on the cell input. Section \ref{II}C is devoted to the study of distortion of response
functions during the storage due to the thermal motion of atoms. Following response functions allows us to estimate numerically the mobility of atoms
calculating the overlap integrals of excited spin modes at the beginning and at the end of storage. In Section \ref{III} we consider another
configuration of the experiment associated with the mixing of atoms in the cell at room temperature. In Section \ref{IV} we optimize the full cycle
of the memory taking into account the thermal motion of atoms at the storage stage.

\section{Quantum memory with slow expansion of atoms}\L{II}

\subsection{Model}\L{model}

Let us begin our research with the quantum memory protocol, based on the resonant interaction of a homogeneous ensemble of three-level atoms in the
$\Lambda$-configuration with signal $\hat E_s$ and driving $E_d$ light pulses. Atoms are situated inside a plane layer with length $L$, which is
orthogonal to $z$-axis. We consider signal and driving light fields as plane waves propagating parallel to $z$-axis. We assume that the driving pulse
is a strong classical field, and the signal pulse is a weak quantum field.

\begin{figure}
\includegraphics[height=7.0cm]{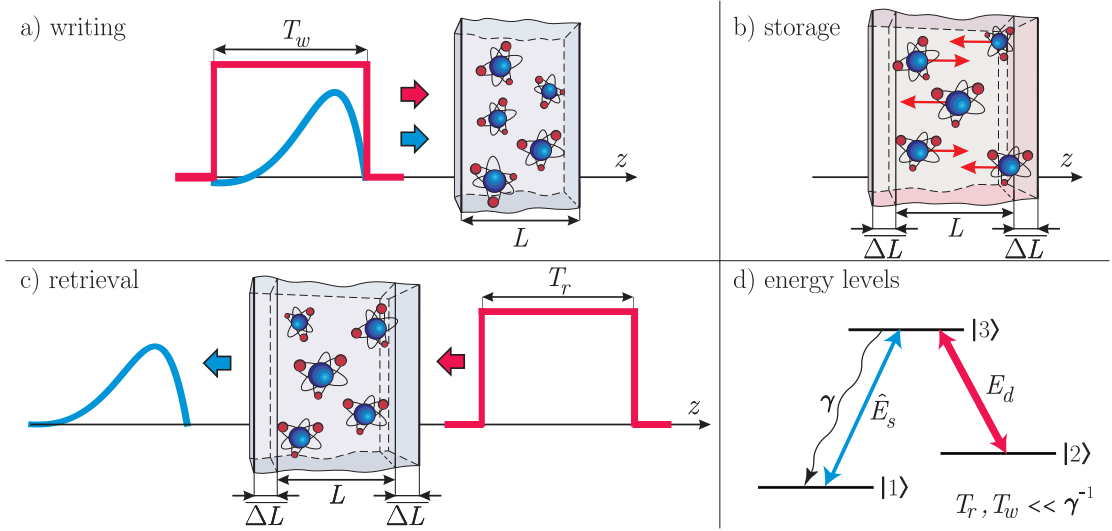}
\caption{Sketch of the full cycle of high-speed quantum memory: a) writing, b) storage with thermal motion, c) retrieval, d) atomic energy levels
with signal $\hat E_s$ and driving $E_d$ light fields.}\L{Fig1}
\end{figure}

Fig.\ref{Fig1} shows stages of the full memory cycle: writing at $0\leq t\leq T_w$ , storage at $T_w<t<(T_w+T_s)$ and retrieval at $(T_w+T_s)\leq
t\leq(T_w+T_s+T_r)$, where $T_w$, $T_s$ и $T_r$ are durations of the each stage, correspondingly.

We choose the durations of the signal and driving pulses much shorter than the time of spontaneous relaxation of the excited state. In particular,
such assumption allows us to take into account the spontaneous decay only during the storage. This choice of interaction times corresponds to so
called high-speed memory regime \cite{HSQM1,HSQM2}.

Before the memory process, the ensemble of atoms is prepared in the ground state  $|1\rangle$. During the writing stage in two-photon interaction the
weak signal field transfer a part of atoms from the state $|1\rangle$ to the excited state $|3\rangle$, and at the same time the strong driving
field, acting simultaneously on the supplementary transition, transfer these atoms from the state $|3\rangle $ to the ground state $|2\rangle $.
Thus, the coherence between levels $|1\rangle$ and $|2\rangle$, which carries all quantum-statistical properties of the signal light pulse, is built.

The ideal storage implies, that the coherence between levels $|1\rangle$ and $|2\rangle$ remains unchanged. However, we will consider two factors of
its distortion: thermal motion of atoms, and spontaneous decay of the residual population from the excited level $|3\rangle$ to the ground level
$|1\rangle$.

At the retrieval stage atoms from the level $|2\rangle$ under the action of the driving pulse return to the level $|1\rangle$ through the excited
level $|3\rangle$. As a result the emission of an output signal occurs, and this pulse completely reproduce the quantum state of the initial signal
or, in the imperfect case, carry some of its properties.

We suppose the interaction processes between the atomic ensemble and the light pulses are short enough, so displacement of atoms during these times
are negligible and we shall consider an atomic thermal motion only during the storage.

In the dipole approximation the light-matter interaction on writing and readout stages is determined by the following Hamiltonian:
\BY
\hat V&&=-\sum_j \hat d_j(t)\hat E(z_j,t),\L{1}\\
\hat E (z_j,t)&&= \hat E_s(z_j,t) + \hat E_d(z_j,t),\nn
\EY
where $\hat d_j$ is the electric dipole operator of the $j$th atom, located in $z_{j}$ at $t=0$.

The signal and driving fields are given as follows (in the plane wave approximation):
\BY &&\hat E_s(z,t)=-i\sqrt{\frac{\hbar\omega_s}{2\varepsilon_0 c S}}e^{\ds -i\omega_s t+ik_{s}z}\hat
a(z,t)+h.c.\;,\\
&&E_d(z,t)=-i\sqrt{\frac{\hbar\omega_d}{2\varepsilon_0 c S}}\;e^{\ds -i\omega_d t+ik_{d}z}\alpha+c.c.\;,
\EY
where $k_{s}$ and $k_{d}$ are wavenumbers of the signal and driving fields. The amplitudes $\hat a(z,t)$ of the signal field and $\alpha$ of the
driving field are normalized so that $\langle\hat a^\dag(z,t)\hat a(z,t)\rangle$ and $|\alpha|^2$ give the mean photon fluxes in photons per second
for the light beam of area $S$. The operators $\hat a(z,t)$ and $\hat a^\dag(z,t)$ obey the commutation relations:
\BY
\[\hat a(z,t),\hat
a^\dag(z,t^\prime)\]&&=\delta(t-t^\prime),\\
\[\hat a(z,t),\hat a^\dag(z^\prime,t)\]&&=c\(1-\frac{i}{k_s}\frac{\partial}{\partial
z}\)\delta(z-z'). \EY
It is natural to describe the light-matter interaction in terms of collective operators, which can be defined as a superposition of the microscopic
variables:
\BY
\hat \sigma_{mn} (z,t)&&=\sum_i|m\rangle\langle n|_i\delta\(z -z_i\),\\
\hat N_m (z,t)&&= \sum_i|m\rangle\langle m|_i\delta\(z -z_i\),\\
\[\hat \sigma_{mn} (z,t),\hat \sigma_{nm} (z',t)\]&&=\( \hat N_m (z,t) - \hat N_n (z,t) \)\delta\(z -z'\).
\EY
Such description allows us to use a continuous variable associated with the observation point instead of discrete coordinates of atom positions.
Although the collective operators turns out to be sufficiently singular, it was shown in \cite{HSQM1} how to perform its smoothing.

Let us rewrite the Hamiltonian in the collective variables:
\BY
&&\hat V=\int dz \;\Big[i\hbar\big(g\hat a(z,t)\hat \sigma_{31} (z,t)\;e^{\ds i k_{s}z}-h.c.) +i\hbar\big(\Omega \hat\sigma_{32}(z,t)\;e^{\ds i k_{d}
z}
-h.c.\big)\Big].%
\EY

Coupling between the light pulses and the atomic ensemble is defined by the coupling constant $g$ and the Rabi frequency $\Omega$, which we choose
real for simplicity:
\BY
&& g=\(\frac{ \omega_s}{2\epsilon_0\hbar c S}\)^{1/2} d_{31},\qquad\Omega=\alpha\(\frac{ \omega_d}{2\epsilon_0\hbar c S}\)^{1/2} d_{32}.
\EY
Here $d_{31}$ and $d_{32}$ are the respective matrix elements of the electric dipole operator.

It is important to notice, that the probability of absorption of signal photons on the transition $|1\rangle-|3\rangle$ is determined not by its
spectral width $\gamma$, but by the product of $\gamma$ and the optical depth $d$. That is why, in spite of the small probability of absorption by a
single atom, the probability of absorption by the whole ensemble is high.

Since the mean number of photons in the signal pulse is small, we consider the population of the level $|1\rangle$ constant during the full memory
cycle and treat it as a real number $N$ \cite{HSQM1, HSQM2}. Taking this into account, it is convenient to re-normalize the coherences and introduce
new atomic operators $\hat b (z,t)$ and $\hat c(z,t)$:
\BY
&& \hat b (z,t)=\hat \sigma_{12} (z,t)/\sqrt{N},\\
&& \hat c (z,t)=\hat \sigma_{13} (z,t)/\sqrt{N},
\EY
which satisfy the bosonic commutation relations. For further calculations we shall derive the simultaneous commutators:
\BY &&\[\hat b(z,t),\hat
b^\dag(z^\prime,t)\]=\delta(z-z^\prime),\\
&&\[\hat c(z,t),\hat c^\dag(z^\prime,t)\]=\delta(z-z^\prime)
\EY

Rewriting the Hamiltonian with the new operators, we define a new coupling constant $g_N$, which is the product of $g$ and $\sqrt{N}$:
$g_N=\sqrt{N}g$. This form of the coupling constant emphasize, that the light-matter interaction is determined by the number of atoms in the
ensemble, and collective effects become apparent for large $N$ resulting in strong interaction.

Let us derive a closed set of Heisenberg equations for operators $\hat a(z,t)$, $\hat b(z,t)$, and $\hat{c}(z,t)$ that describes the evolution on the
writing and the retrieval stages, i.e. at $0\leq t \leq T_w$ and $(T_w+T_s)\leq t \leq(T_w+T_s+T_r)$:
\BY
  \frac{\partial}{\partial
z}\hat a(z,t)&&=- g_{N}\;
\hat c(z,t),\L{1.12}\\
\frac{\partial}{\partial t}{\hat c}(z,t)&&=
     g_{N}\;\hat a(z,t)+\Omega\;
 \hat b(z,t),\\
  \frac{\partial}{\partial t}{\hat b}(z,t)&&= -
 \Omega\hat c(z,t).\L{1.14}
\EY
The derivation of the equations and the way of their solution can be find in \cite{HSQM1, HSQM2}, where the protocol of high-speed quantum memory on
cold atoms were proposed. Solutions (in dimensionless variables) connect the amplitude of the signal field $\hat a_{in}(\tilde t)$ on the input of
the cell with the coherence $\hat b(\tilde z,\tilde T_w)$ at the end of writing,
 \BY
 && \hat b(\tilde z,\tilde T_w)= -\int_0^{\tilde T_w} d{\tilde t} \; G_{ab}(\tilde z,\tilde t)\hat a_{in}(\tilde t)+vac,\L{18}\\
 && G_{ab}(\tilde z,\tilde t)= \frac{1}{\sqrt{2}}\int_0^{\tilde t} d{\tilde t'}g_{ab}(\tilde z,\tilde t')g^{\ast}_{ab}(\tilde z,\tilde t-\tilde t'),
 \qquad g_{ab}(\tilde z,\tilde t)=e^{-i\tilde t}J_0\(\sqrt{\tilde z \tilde t}\)\Theta_w(\tilde t),\nn
\EY
as well as the coherence  $\hat b( z,T_w+T_s)$ at the end of the storage with the output field $\hat a_{out}(t)$ obtained in backward retrieval
\BY
 && \hat a_{out}(\tilde t)= -\int_0^{\tilde L} d{\tilde z} \; G_{ba} (\tilde z,\tilde t)\hat b(\tilde z,\tilde T_w+\tilde T_s)+vac,\L{19}\\
 && G_{ba}(\tilde z,\tilde t)= \frac{1}{\sqrt{2}}\int_0^{\tilde t}  d{\tilde t'}g_{ba}(\tilde z,\tilde t')g^{\ast}_{ba}(\tilde z,\tilde t-\tilde t'),
 \qquad g_{ba}(\tilde z,\tilde t)=e^{-i\tilde t}J_0\(\sqrt{\tilde z \tilde t}\)\Theta_r(\tilde t),\nn
\EY
Here $J_0$ is the zero order Bessel function of the first kind, $\Theta_w (\tilde t)$,  $\Theta_r (\tilde t)$ are the window-functions (they are
equal to one during the writing and the retrieval stages, correspondingly, and zero-valued for all the other time intervals). Dimensionless
coordinate $\tilde z$ and time $\tilde t$ defined as follows:
\BY
&& \tilde z = \frac{2g_N^2 }{\Omega}z\qquad \tilde t = t \Omega,\L{1.20}
\EY

Henceforth we will omit "tilde"\; over variables, regarding them dimensionless, if the otherwise not stated.

Note, that in Eqs. (\ref{18})-(\ref{19}) there are terms, which marked as $vac$ and correspond to contributions of the subsystems in the vacuum
state. There is no need to specify them, since their contributions vanish when normally ordered expectation values are taken.

To consider atomic motion during the storage and estimate its influence on the stored coherence $\hat b(z, T_w)$, let us introduce a subensemble of
atoms moving co-directionally with a velocity $v_z$. Then evolution of the coherence $\hat b(z, T_w; v_z)$ of this subensemble during the storage is
determined by the following equation
\BY
 && \(\frac{\partial}{\partial t}+v_z\frac{\partial}{\partial z}\){\hat b}(z,t; v_z)=0.\L{1.21}
\EY
The transition from the ensemble to the subensemble could be formally done as far back as in the Hamiltonian (\ref{1}), if one redefine properly the
collective variables $\hat b(z, t)$ and $\hat c(z, t)$ for the subensemble. However, we prefer to perform it now to underline, that we consider
atomic motion only during the storage. Note, the equations (\ref{1.12})-(\ref{1.14}) and their solutions (\ref{18})-(\ref{19}) are valid not only for
the ensemble of motionless atoms, but also for any its subensemble, which moves as a whole with some certain velocity.

The solution of Eq. (\ref{1.21}) at the end of the storage $T_w+T_s$ can be got in a simple form:
\BY
&&\hat b(z, T_w + T_s; v_z) = \hat b(z - v_z  T_s, T_w; v_z),\L{1.22}
\EY
where $v_z$ is a dimensionless velocity defined according (\ref{1.20}). The equation (\ref{1.22}) shows that at the end of the storage stage the
domain of function $\hat b(z, T_w+ T_s;  v_z)$ is shifted on $v_z T_s$, from $ z\in\[0, L\]$ at the beginning of the memory cycle to $z\in\[v_z T_s,
L+ v_z T_s\]$ at the end of the storage. At the same time, the profile of the function remains unchangeable. Thereby, to describe the motion of all
atoms, we shall take into account all contributions $\hat b(z, T_w+ T_s; v_z)$ from each subensemble to the coherence $\hat b(z, T_w+ T_s)$ of the
whole ensemble, and this coherence is the initial condition for the retrieval stage.

Further analysis of the solutions derived in this Section we will carry out in terms of eigenfunctions and eigenvalues of the memory cycle.

\subsection{Eigenfunctions of the memory cycle and response functions}

Let us start with the consideration of cold atoms, which stay immobilized during the full memory cycle. The connection between the input and the
output (retrieved) fields is described by the integral transformation:
\BY
&& \hat a_{out} (t) = \int\limits^{T_w}_{0}dt'\; \hat a_{in} (T_w-t') G(t,t')+vac.\L{25}
\EY
Here $G(t,t')$ is the real kernel of the integral operator. It can be expressed by the kernel of the writing stage $G_{ab}(z,t)$, converted the input
field $\hat a_{in}(t)$ to the spin coherence $\hat b(z, T_w)$, and the kernel of the retrieval stage $G_{ba}(z,t)$, converted the spin coherence
$\hat b(z, T_w)$ to the output field $\hat a_{out}(t)$:
\BY
&& G(t,t')=\int\limits_{0}^{L}dz\;G_{ab}(z,t)G_{ba}(z,t').\L{26}
\EY
Since in the high-speed memory regime $G_{ab}(z,t) = G_{ba}(z,t)$, than the kernel $G(t,t')$ is symmetric with respect to permutation of the
arguments $t \leftrightarrow t'$. This means we have a right to derive the equation for its eigenfunctions $\{\phi_{i}(t)\}_{i=1}^{\infty}$ and
eigenvalues $\{\sqrt{\lambda_{i}}\}_{i=1}^{\infty}$ in the form:
\BY
&& \sqrt{\lambda_i} \phi_i (t) =\int\limits_{0}^{T_w}dt'\;G(t,t')\phi_i(t').\L{27}
\EY
The functions $\{\phi_{i}(t)\}_{i=1}^{\infty}$ form the complete orthonormal set:
\BY
&& \int\limits_{0}^{T_w} dt\; \phi_{i}(t)\phi_{j}(t)=\delta_{ij},\L{28}\\
&& \sum\limits_{i=1}^{\infty}\phi_{i}(t)\phi_{i}(t')=\delta(t-t'). \L{29}
\EY

One should note, that such definition of the eigenfunctions assumes the equality of the writing and the retrieval durations. In more general case,
the duration of the retrieval can exceed the writing time, so the arguments of the kernel $G(t,t')$ will be defined on the different domains.
However, even in this case it is possible to symmetrize the kernel and find its eigenfunctions and eigenvalues (see Appendix A).

One can derive the equation equivalent to the formula (\ref{27}), representing the kernel $G(t,t')$ as a bilinear quadratic form of its
eigenfunctions, where the corresponding eigenvalues are decomposition coefficients:
\BY
G(t,t')=\sum\limits_{i=1}^{\infty}\sqrt{\lambda_{i}}\;\phi_{i}(t)\phi_{i}(t'). \L{30}
\EY
Such representation commonly called the Schmidt decomposition, and $\phi_{i}$ are referred to as the Schmidt modes.
\begin{figure}[t]
\includegraphics[height=6.0cm]{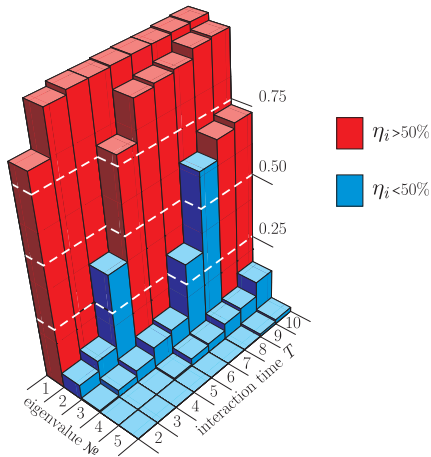}
\caption{The first five eigenvalues as a function of the duration $T$ of the writing and the retrieval processes, when $T_r=T_w=T$ and $L=10$. The
red columns correspond to the eigenfunctions with efficiency $\eta_i>50\%$, the blue columns -- with efficiency $\eta_i<50\%$.}
\L{Fig2}
\end{figure}

The diagram in Fig. \ref{Fig2} depicts the first five eigenvalues as a function of the duration of the writing and the retrieval light pulses, when
$T_r=T_w=T$ and the length of the atomic layer is $L=10$. It shows the eigenfunctions with the efficiency higher then $50$\% (red columns) that means
quantum memory regime, and below it (blue columns). In other words, if the input signal profile coincides with the $i$th eigenfunction, than the
quantum memory performs like a beamsplitter with the transmission coefficient $\eta_i$, which equals to the squared $i$th eigenvalue and determines
the efficiency of the memory:
\BY
&& \eta_i =\frac{\int_0^T dt \langle \hat a_{out}^\dag(t) \hat a_{out}(t) \rangle}{\int_0^T dt \langle \hat a_{in}^\dag(t) \hat a_{in}(t)
\rangle}=\lambda_i.\L{1.29}
\EY
Let us make a remark, that if a time profile of the input pulse is not identical to any of eigenfunction, then the description of the protocol as a
beamsplitter with a certain coefficient is not valid longer. The properties of such protocol will be determined by the set of numbers associated with
projections of the input temporary profile on the eigenfunctions of the memory \cite{HS2014}.

As one can see from the picture, for any  $T$ eigenvalues decrease rapidly and, in fact, only the first two are noticeably different from zero. That
is why we will consider further only the first two eigenfunctions.

Note, this memory model is valid only when $T<L\;$\cite{HSQM1}.

Further we will consider only the case of backward retrieval at $L=10$ and $T=5.5$. We exploited such values of parameters in \cite{HS2014} to find
eigenvalues and eigenfunctions in the case of motionless atoms, and now we want to compare them with the case of thermal motion.

Fig. \ref{Fig3} shows the first two eigenfunctions of the full memory cycle for motionless atoms (left column) and their squares (right column).

\begin{figure}[t]
\includegraphics[height=7.0cm]{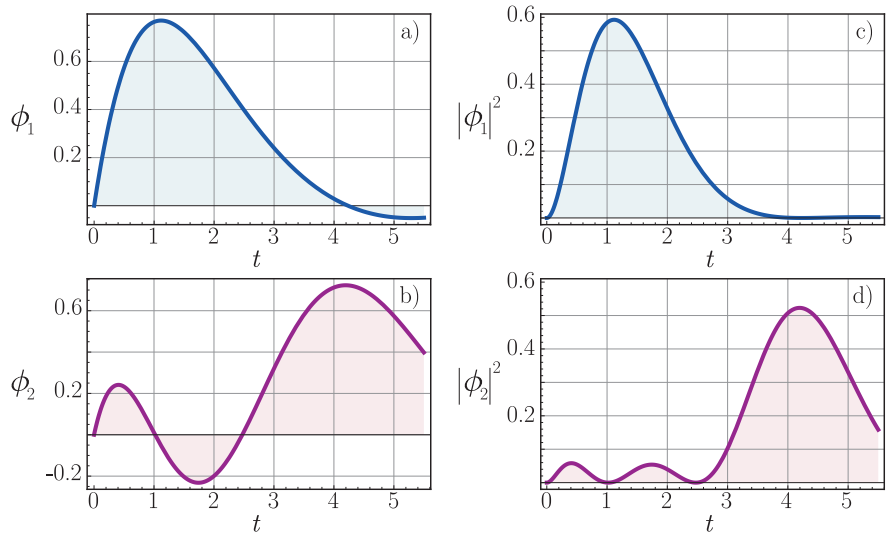}
\caption{The eigenfunctions $\phi_1(t)$ and $\phi_2(t)$ of the memory cycle for motionless atoms (left column, frames a and b), and their squares
(right column, frames c and d). Dimensionless time $t$ is given in units of $\Omega^{-1}$.} \L{Fig3}
\end{figure}

Let us now consider a situation, when the signal with temporary profile of one of the memory eigenfunctions incidents on the input face of the atomic
cell, and find out, what kind of "response" of the medium would caused by such a field. We shall call this transformation the "half-cycle" as the
opposite to the full memory cycle of writing and retrieval of the signal:
\BY
&& \sqrt{\mu_i} \psi_{i}(z)=\int\limits_{0}^{T_w}dt\;G_{ab}(t,z)\phi_{i}(t).\L{31}
\EY

As will be proved below, the set $\psi_{i}(z)$ is orthonormal and the normalization factor $\sqrt{\mu_i}$ is related with the eigenvalues $\lambda_i$
of the full cycle. We shall call products $\sqrt{\mu_i} \psi_{i}(z)$  the \emph{response functions} of the medium.

In the contrast to the kernel $G(t,t')$ of the full memory cycle, the kernel $G_{ab}(t,z)$ is not symmetric in respect to the permutation of its
temporary argument $t$ and spatial argument $z$, and so the Schmidt decomposition is ineligible here. However, due to the completeness of the
eigenfunctions $\{\phi_{i}(t)\}_{i=1}^{\infty}$, we can expand $G_{ab}(t,z)$ into the series:
\BY
&& G_{ab}(t,z) = \sum_{i=1}^\infty g_i(z) \phi_i(t),\L{1.31}
\EY
here $g_i(z)$ are the expansion coefficients. Now we can find the relation between $g_i(z)$ and $\sqrt{\mu_i} \psi_{i}(z)$. For this we multiply the
right and the left parts of Eq. (\ref{1.31}) on $\phi_j(t)$ and then integrate this expression by time from $0$ to $T_w$. Taking into account the
orthonormality of $\{\phi_i(t)\}_{i=1}^\infty$, we get
\BY
&& \sqrt{\mu_j} \psi_{j}(z)=g_j(z).
\EY
This relation means, that the $i$th response function of the medium  $\sqrt{\mu_i} \psi_{i}(z)$ is the $i$th expansion coefficient of the
$G_{ab}(t,z)$ by the eigenfunctions $\{\phi_i(t)\}_{i=1}^\infty$, so the equation (\ref{1.31}) can be rewritten in form
\BY
&& G_{ab}(t,z) = \sum_{i=1}^\infty \sqrt{\mu_i} \psi_{i}(z) \phi_i(t).
\EY
Now let us prove, that functions $\psi_{i}(z)$ compose a complete orthonormal set, and find the normalization factors $\sqrt{\mu_i}$. Express the
scalar product of the $i$th and the $j$th response functions, using Eq. (\ref{31}):
\BY
&& \sqrt{\mu_i}\sqrt{\mu_j}\int_0^L\;dz \psi_{i}(z) \psi_{j}(z) =\int_0^{L} \int_0^{T_w} \int_0^{T_w}dzdtdt'\;
G_{ab}(t',z)G_{ab}(t,z)\phi_i(t')\phi_j(t)
\EY
Taking into account the Eqs. (\ref{28})-(\ref{30}), we obtain
\BY
&& \sqrt{\mu_i}\sqrt{\mu_j}\int_0^L\;dz \psi_{i}(z) \psi_{j}(z) = \sqrt{\lambda_i}\delta_{ij}.
\EY
This means that functions $\psi_{i}(z)$ are orthonormal, with the normalization factors $\sqrt{\mu_i}=\sqrt[4]{\lambda_i}$. As a result the following
expansion for $G_{ab}(t,z)$ can be written:
\BY
&& G_{ab}(t,z) = \sum_{i=1}^\infty \sqrt[4]{\lambda_i} \psi_{i}(z) \phi_i(t).\L{1.36}
\EY
The completeness of the set $\{\psi_{i}(z)\}_{i=1}^\infty$ is followed from the linearity of the integral transform (\ref{31}) and the completeness
of the set $\{\phi_{i}(t)\}_{i=1}^\infty$.

The functions  $\psi_{i}(z)$ can be conditionally called the eigenmodes of the spin system. Further, we will be mostly interested not in these
eigenmodes, but in the corresponding response functions $\sqrt[4]{\lambda_i}\psi_{i}(z)$, squares of which reveal the distribution of the excitations
within the medium.
\begin{figure}[t]
\includegraphics[height=7.0cm]{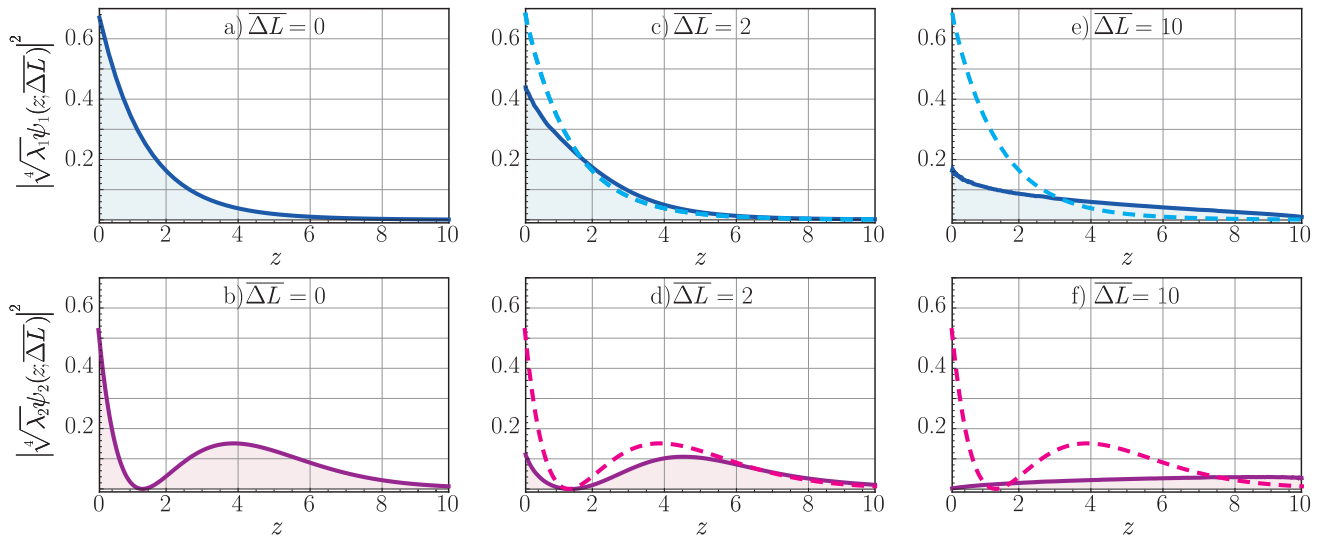}
\caption{Squares of the response functions. Top row -- $|\sqrt[4]{\lambda_1}\psi_1(z;\overline{\Delta L})|^2$, bottom row --
$|\sqrt[4]{\lambda_2}\psi_2(z;\overline{\Delta L})|^2$. In frames a) and b) $\overline{\Delta L}=0$ (motionless atoms), in frames c) and d) --
$\overline{\Delta L}=2$, in frames e) and f) -- $\overline{\Delta L}=10$. Dashed curves are given for ease of comparison and correspond to
$\overline{\Delta L}=0$. Dimensionless coordinate $z$ is given in units of the effective optical depth (see subsection \ref{scaling}).} \L{Fig4}
\end{figure}

Figure \ref{Fig4}a shows the spatial dependence of the first and second response functions for the motionless atoms. The area under the curve for the
first response function is $\sqrt{\lambda_1}=1.0$, for the second one -- $\sqrt{\lambda_2}=0.8$ that is well agreed with the corresponding storage
efficiencies. Besides, one can take notice, that the first curve has only one peak at $z=0$, at the same time the second curve has two peaks at $z=0$
and at $z\approx3.9$. Thus, in the first case the excitations are localized mainly near the input face of the cell, whereas in the second case they
partially "pushed" in the middle. In the next Section we will follow how the obtained curves changes due to the thermal motion at the storage stage.

\subsection{Thermal motion at the storage stage for slow atoms}

\subsubsection{The limits of applicability}

Now let us consider the atomic motion in the longitudinal direction at the storage stage, when $T_w<t<T_w+T_s$. Well known, that an ideal gas can be
described in the framework of classical statistics, when its temperature exceeds the degeneracy temperature \cite{LandauLifshitz}:
\BY
&&T\gg\frac{n^{\frac{2}{3}}h^2}{3mk}, \L{1.37}
\EY
here $T$  is a temperature, $n$ is a volume concentration of the atoms, $m$ is a mass of a single particle, $k$ -- Boltzmann constant, $h$ -- Plank
constant. The macroscopic behavior of nondegenerate gas obeys Maxwell-Boltzmann statistics.

We consider the case, when an atomic ensemble has Maxwell speed distribution, because it covers wide class of the quantum memory systems. For
example, it takes place in the works \cite{LauratQM1,LauratQM2}, where authors experimentally consider quantum memory on the caesium vapors with the
average concentration of atoms about $10^6$ particles in $1$ mm$^3$ at the temperature about $100$ $\mu K$. One can verify, that such parameters
satisfy the inequality (\ref{1.37}).

In Section \ref{model} we derived the equation (\ref{1.22}), which bonds a coherence of spin subensemble moving as a whole in the longitudinal
direction with the velocity $v_z$ at the moment $t=T_w$ and a coherence of the same subensemble at the end of the storage $t=T_w+T_s$. Let us rewrite
this expression for the $i$th response function:
\BY
&& \sqrt[4]{\lambda_{i}}\psi_i(z)\Big|_{{v_z},t=T_w+T_s}= \sqrt[4]{\lambda_{i}}\psi_i(z - v_z T_s)\Big|_{{v_z},t=T_w},
\EY
and pass from the subensemble to the whole ensemble with the Maxwell speed distribution:
\BY
&& \sqrt[4]{\lambda_{i}}\psi_i(z)\Big|_{t=T_w+T_s} = \sqrt[4]{\lambda_{i}}\frac{1}{\sqrt{\pi} u_z}\int_{-\infty}^{+\infty}dv_z\;
e^{-\frac{v_z^2}{u_z^2}}\psi_i(z - v_z T_s)\Big|_{{v_z},t=T_w},
\EY
where $u_z$ is the root-mean-square velocity of the particles in the longitudinal direction. The mean extension of the spin ensemble during the
storage is given by $\overline{\Delta L}$:
\BY
&&\overline{\Delta L}= T_s u_z,
\EY
which specify the mean temperature of the ensemble (for the given $T_s$). Hereinafter we will use denotation $\psi_i(z;\overline{\Delta L})$ instead
of $\psi_i(z)\Big|_{t=T_w+T_s}$.

\subsubsection{Scaling of the coordinate}\L{scaling}

One can see that the definition of dimensionless coordinate (\ref{1.20}) for motionless atoms implies the atomic concentration $N$ as constant, then
the dimensionless coordinate coincides with an effective optical depth (which differ from the real optical depth in $\Omega/\gamma$ times). In other
words, in the case of motionless atoms the dimensionless coordinate is measured in units of the effective optical depth. When the dimensional
coordinate varies in $z \in [0, L]$, the dimensionless coordinate varies in $\tilde z \in [0, \tilde L]$. (In this subsection we again return to the
old notations with tilde for the dimensionless variables.)

As a result of thermal motion, the homogeneous distribution of atoms is disturbed, and the concentration of atoms $N$ becomes a function of $z$, with
a domain ${z}\in [-\infty,+\infty]$ (nevertheless, the most of atoms are located within $z \in [-\overline{\Delta L},L+\overline{\Delta L}]$)). This
change breaks the direct correspondence between the dimensionless coordinate $\tilde z$ and optical depth. We want to redefine the dimensionless
coordinate so that to restore this equivalence. We will introduce new dimensionless coordinate $\bar z$, which will vary in the same interval $\bar z
\in [0,\tilde L]$, as the old one. By this reason we carry out the scaling of the coordinate.

The main idea of the scaling is to find the relation $\bar z=f(\tilde z)$ between the old dimensionless coordinate $\tilde z$ and a new one $\bar z$
such that concentration $\bar N(\bar z)$ is again homogeneous.

As far as the longitudinal motion does not change the whole number of atoms in the system, we can always write the following equality
\BY
    && \int_{0}^{\tilde L}d\tilde z\; N(\tilde z)= \int_{0}^{\bar{L}}d\bar{z}\; \bar{N}(\bar{z}).
\EY
Here the left part corresponds to the whole number of atoms before storage, and $N(\tilde z)$ is a constant. The right part is the same number after
the thermal motion, $\bar N$ can be obtained from $N(\tilde z)$  by averaging it over the Maxwell distribution, and we demand $\bar N(\bar z)$ is
also constant. Let us replace in the right part $\bar{z}$ on $\tilde z$: $\bar{z}=f(\tilde z)$, $d\bar{z}=f'(\tilde z)d\tilde z$, and $\bar L=\tilde
L$, we get
\BY
    && \int_{0}^{\tilde L}d\tilde z\; N(\tilde z)= \int_{0}^{\tilde L}d\tilde z\; f'(\tilde z)\bar{N}\big(f(\tilde z)\big).
\EY
From here one can obtain
\BY
    && f'(\tilde z) = \frac{N(\tilde z)}{\bar{N}\big(f(\tilde z)\big)}.
\EY
As far as $N(\tilde z)$ and $\bar{N}\big(f(\tilde z)\big)$ are known, we can solve this equation numerically, and reconstruct the scaling function
$f(\tilde z)$.

\subsubsection{Response functions for mobile atoms}

Thermal motion of atoms leads to the spatial redistribution of the stored spin coherence. Let us now apply the scaling procedure described in the
previous subsection, and follow the influence of such redistribution on the response functions $\sqrt[4]{\lambda_{i}}\psi_i(z;\overline{\Delta L})$.
Hereinafter we again omit "bar"\; over $z$, regarding it as the dimensionless variable expressed in optical depth units (with recipe given above).

Figs. \ref{Fig4}b and \ref{Fig4}c show the spatial dependence of squares of the first and the second response functions for mobile atoms, when the
mean shift of atoms equals $\overline{\Delta L}=2$ (i.e. one fifth of the cell length) and $\overline{\Delta L}=10$ (i.e. the full cell length),
correspondingly. As before, the top row (blue curves)  corresponds to the first response function, the bottom row (purple curves) -- to the second.
For simplicity of comparison, the dashed curves mark the case of motionless atoms. First of all we see, that for small displacements
($\overline{\Delta L}=2$) the shapes of the curves and the areas under the curves have changed slightly comparing with the case of motionless atoms.
So the spatial distribution of excitation in the medium changes insignificantly, and we can expect that after retrieval we receive an output field
with a time profile similar to the corresponding input eigenmode with high efficiency. Note, the both curves have the same number of peaks, as they
had in the case of motionless atoms. However, due to thermal motion these peaks noticeably subsided, and the peak at $z\approx3.9$ for the second
curve shifted to the right in $z\approx4.6$. From the figure one can conclude that the stored excitation in the medium was "blurred" by the thermal
motion over the whole spin ensemble. This is clearly seen in the case of the mean displacement equals $\overline{\Delta L}=10$. Besides, it shows,
that the higher number of zeros of the mode $\psi_i(z)$ and the lower "contrast" of its peaks, the faster "blur" will destroy this mode. In
particular, we see, that at $\overline{\Delta L}=10$ the peak at the beginning of the first response function, in spite of its subsidence, is
preserved, whereas the both peaks of the second response function have vanished.

\subsubsection{Overlap integrals}

We have been convinced, that thermal motion redistribute the stored spin excitations from the initially excited mode over the others spin modes. At
the same time, the main role in the recovery of the signal after such redistribution will play the modes with the highest eigenvalues
$\sqrt{\lambda_i}$. Therefore one need to retrieve the signal not only from the initially excited mode but also from all the others with high
$\sqrt{\lambda_i}$ for the effective retrieval. To confirm such interpretation quantitatively, we introduce the overlap integrals
$Q_{ij}(\overline{\Delta L})$ of the $i$th "blurred"\; mode of the thermal atomic ensemble with the $j$th eigenmode of the immobilized ensemble:

\BY
    &&Q_{ij}(\overline{\Delta L})=\int_{0}^{L}\;dz \psi_i(z;\overline{\Delta L}) \psi_j(z;0),\L{44}
\EY
where $\psi_j(z;0)\equiv\psi_j(z)$. It is seen from the definition that overlap integrals can be rewritten in the matrix form, which has $Q_{ij}$ as
the element of the $i$th row and the $j$th column.

Then, for $\overline{\Delta L}=2$ the overlap integrals are given by
$$Q=\begin{pmatrix} 0.92 &0.11 \\ 0.11 & 0.74 \end{pmatrix},$$
and for  $\overline{\Delta L}=10$ we got
$$Q=\begin{pmatrix} 0.68 & 0.32 \\ 0.32 & 0.39 \end{pmatrix}.$$
These numerical calculations confirm our previous conclusions. The square of the first response function has only one peak, whereas the square of the
second one -- two small peaks. As a result, the second response function will be changed by thermal motion more significantly than the first one,
what can be seen from comparison of $Q_{11}$ and $Q_{22}$. In other words, the first function can be stored better, than the second.  The equal
values of $Q_{12}$ and $Q_{21}$ reflects the number of excitations, which flowed from the first mode to the second one and vice versa. One can
notice, that in the case of $\overline{\Delta L}=0$  (motionless atoms) $Q_{12}$ and $Q_{21}$ are equal to zero. Thus, in any cases the matrix $Q$ is
symmetrical.

One can evaluate the efficiency of the full memory cycle over the overlap integrals. For this purpose let us expand the "blurred"\; response function
$\sqrt[4]{\lambda_i}\psi_i(z;\overline{\Delta L})$ by the full orthonormal set $\{\psi_j(z)\}_{j=1}^{\infty}$. Then, taking into account Eq.
(\ref{44}) we get
\BY
    &&\sqrt[4]{\lambda_i}\psi_i(z;\overline{\Delta L})=\sqrt[4]{\lambda_{i}} \sum_{j=1}^{\infty}Q_{ij}\psi_{j}(z).
\EY
Using this formula, Eq. (\ref{19}), and Eq. (\ref{1.36}), we can derive the time profile $\phi_i^{out}(t)$ of the output signal at the readout, when
the incident light pulse had the profile of the $i$th eigenfunction:
\BY
    &&\phi^{out}_i(t)=\sum_{j=1}^{\infty}Q_{ij}\sqrt[4]{\lambda_{i}\lambda_{j}}\phi_{j}(t).\L{46}
\EY

Substituting Eq. (\ref{46}) into the definition of the memory efficiency (\ref{1.29}), we got
\BY
    &&\eta_i=\sum_{j=1}^{\infty}Q_{ij}^2\sqrt{\lambda_{i}\lambda_{j}}.\L{1.49}
\EY
Using this formula, we obtain the following values of efficiency. In the case of the mean displacement $\overline{\Delta L}=2$ for the input pulse
shaped like $\phi_1(t)$, the efficiency of the full memory cycle is $\eta_1=87\%$, like $\phi_2(t)$ -- the efficiency $\eta_2=46\%$. In the case of
$\overline{\Delta L}=10$ we obtained $\eta_1=55\%$ and $\eta_2=22\%$. All these values completely coincide with the results of direct numerical
calculations of $\eta_i$, based on the initial integral transformations. As one can see, even for relatively large temperatures of atoms and without
any additional optimizations, one can choose the time profiles, which provide the quantum level of the efficiency (higher than $50\%$ ).

Note, the diagonal elements of $Q$ correspond to single-mode efficiency (so called beamsplitter-efficiency) of quantum memory, when one readout the
signal from the same eigenmode, in which it was written. With respect to this mode, quantum memory works as a beamsplitter with transmission
coefficient equals to storage efficiency. Single-mode efficiency is remarkable because it describes not only the quality of storage of the photon
numbers, but also all the other moments of quantum state distribution of the field. It is important to remember that this is the only case where the
efficiency is a universal characteristic of the system. If the recorded signal is not an eigenfunction of memory, or reading mode does not match the
writing mode, the relationship between efficiency and preservation of other quantum properties of the field becomes more complicated
\cite{Golubev2015}.
\begin{figure}[t!]
\includegraphics[height=7.0cm]{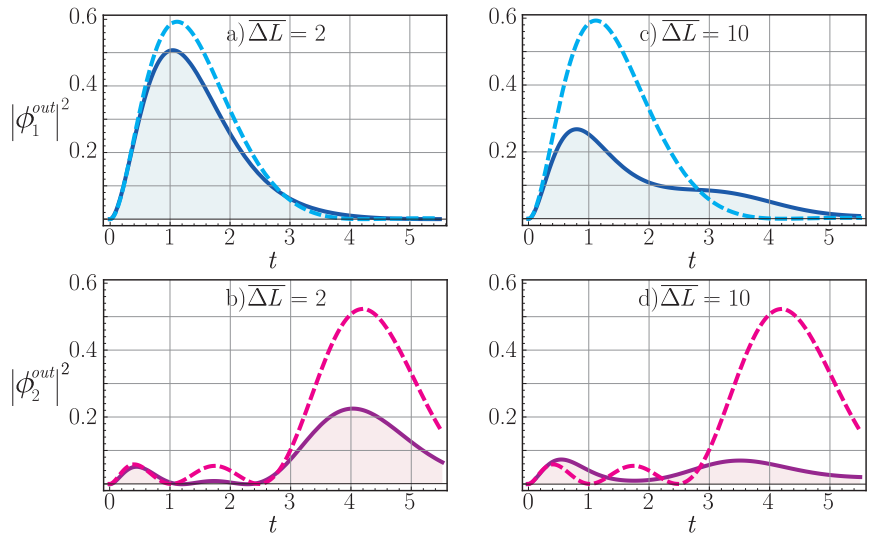}
\caption{The time profiles of the intensity of the output fields. Top row -- $|\phi^{out}_1(t)|^2$; bottom row -- $|\phi^{out}_2(t)|^2$. In frames a)
and b) $\overline{\Delta L}=2$, in frames c) and d)  $\overline{\Delta L}=10$. Dashed curves are given for ease of comparison and correspond to
$\overline{\Delta L}=0$. Dimensionless time $t$ is given in units of $\Omega^{-1}$.} \L{Fig5}
\end{figure}

In Fig. \ref{Fig5} the squares of $\phi^{out}_i(t)$ are plotted. As before, the top row (blue curves)  corresponds to the first eigenfunction, the
bottom row (purple curves) -- to the second. Dashed curves correspond to $\overline{\Delta L}=0$  and given for comparison. It is interesting to
notice that the redistribution of excitation over the full memory cycle eigenfunctions looks more visual, comparing with the same process for the
response functions (see Fig. \ref{Fig4}). In particular, one can see the second peak, specific for $\phi^{out}_2(t)$, which grows on
$|\phi^{out}_1(t)|^2$ at $\overline{\Delta L}=10$.

\section{Storage in the atomic cell at room temperature}\L{III}

Let us now consider the atomic ensemble at room temperature. In this case we can not longer discuss mean displacements of atoms of the order of cell
length, as it was in the previous Section. We will suppose that at room temperature the full mixing of the atoms in the cell during the storage time
occurs, and at the end of the storage the distribution of spin coherence becomes homogeneous.

As opposite to the  model discussed before, we will suppose now that the atomic ensemble is situated not in a free space, but in the closed prolate
cell with length $L$ (along the signal and the driving pulses propagation), and with the narrow cross-section (with no transverse degrees of
freedom). As before, we will study the regime of the high-speed quantum memory and treat atoms as immobilized at the writing and readout stages, due
to the shortness of the interaction times.

We imply that cell coated with the spin protecting alkene coating \cite{Balabas1, Balabas2, Balabas3} and any atomic relaxations caused by collisions
with walls can be eliminated.

Because now atoms do not leave the cell, and their concentration stays uniform, the definition of the dimensionless coordinate remains similar to the
case of motionless atoms, and we do not need to apply any scaling.

We shall emphasize, that in contrast to immobilized atoms, any spin distribution, formed in the writing process, will be blurred by the thermal
motion and converted into uniform, independently of the shape of input pulse. However, the input pulse shape will determine the total number of spin
excitations. As a result, the output signal shape will be always the same, and only difference will be revealed in the efficiency of the full memory
cycle. The latter, as before, can be calculated with the help of the overlap integrals (\ref{1.49}).

Thus, in order to analyze this problem numerically, we should examine only the retrieval, assuming that the spatial distribution of the spin
coherence at the end of storage stage is known (this is the constant function, which level corresponds to the total number of spin excitations). We
have confined ourselves to the case of the input field profile matched with $\phi_1(t)$.

\begin{figure}
\includegraphics[height=3.5cm]{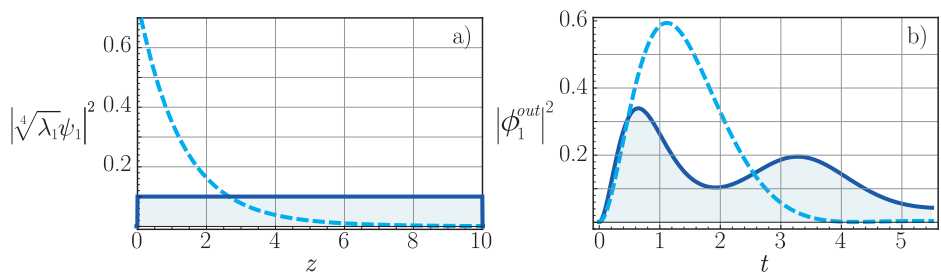}
\caption{a) The square of the first response function right after the writing (dashed curve) and after the storage (solid curve). b) The time profile
of the output field intensity stored on room temperature atoms (solid curve) and motionless atoms (dashed curve). Dimensionless time $t$ is given in
units of $\Omega^{-1}$. Dimensionless coordinate $z$ is given in units of effective optical depth (see subsection \ref{scaling}).} \L{Fig6}
\end{figure}
The dashed curve in the Fig. \ref{Fig6}a shows the square of the response function $\sqrt{\lambda_1}\psi_1(z)$ right after the writing stage. The
solid curve on this plot corresponds to the same function, but after the storage time. In the Fig. \ref{Fig6}b the solid curve depicts the intensity
profile of the output field, and the dashed line is given for comparison with the case of motionless atoms. One can see, that complete mixing of
atoms inside the cell leads to redistribution of the retrieved intensity and to formation of two peaks at $ t\approx 0.6$ and $t\approx 3.4$, which
are characteristic for the first and the second eigenfunctions. The first peak is higher than the second one, since the efficiency for the first mode
is higher.

The numerical calculations bring out the high values of the overlap integrals for room temperature atoms: $Q_{11}=0.73$ and $Q_{12}=0.56$ that
provide the total memory efficiency $\eta_1$=82\%. Thereby, complete mixing of atoms inside the cell, contrary to our expectations, does not lead to
the flagrant damage of the memory, comparing with the case of slow atoms in a free space considered before. In particular, at $\overline{\Delta
L}=10$ we obtained the worse results for slow atoms in a free space.

Thus, we can conclude that, for the given optical depth, the choice of a particular experimental configuration may lead to improvement or
deterioration of the results depending on the overlap of the stored spin distribution with the most efficient spin eigenmodes.

\section{Optimization of full memory cycle}\L{IV}

In this article, to ensure the maximum efficiency of the memory protocol, we applied the optimization based on the analysis of the kernel of integral
transformation and fitting the optimal signal profile. This approach was introduced as a tool of the quantum memory optimization in \cite{Nunn2007}.
(Note, that the other well-known optimization method based on driving field shaping \cite{Gorshkov1, Gorshkov2}, remains outside of this discussion.)
It is clear that thermal motion of atoms at the storage stage leads to change of the kernel of integral transformation: the eigenfunctions for the
case of immobilized atoms will no longer be the eigenfunctions of the task with mobile atoms. Thus, it is natural to analyze the new integral
transformation taking into account the motion of the atoms, and specify which profile of incident pulse ensures the maximum efficiency.

Let us deduce a new kernel of the full memory cycle $G(t,t';\overline{\Delta L})$ instead of Eq. (\ref{26}). Previously, both factors $G_{ab}(t,z)$
and $G_{ba}(t,z)$, which describe singly the writing and readout stages, were identical; now we include into the first of them the atomic motion
during the storage. Formally, this brings to replacement of the response functions by the average response function $\sqrt[4]{\lambda_i}
\psi_i(z;\overline{\Delta L})$ into the expansion of $G_{ab}(t,z)$. Consequently, the new kernel for the writing stage $G_{ab}(t,z; \overline{\Delta
L})$ is given by
\BY
&& G_{ab}(t,z;\overline{\Delta L}) = \sum_{i=1}^\infty \sqrt[4]{\lambda_i} \psi_i(z;\overline{\Delta L}) \phi_i(t).
\EY
The kernel of the integral transformation for the reading stage $G_{ba}(t,z)$ remains unchanged. As a result, the new kernel for the full memory
cycle $G(t,t'; \overline{\Delta L})$ is following:
\BY
&& G(t,t';\overline{\Delta L})=\int\limits_{0}^{L}dz\;G_{ab}(t,z;\overline{\Delta L})G_{ba}(t',z),\L{1.50}
\EY
Developing the kernels $G_{ab}(t,z;\overline{\Delta L})$ and $G_{ba}(t',z)$ as series in eigenfunctions $\phi_i(t)$, one can be certain that they
remain symmetrical with respect to permutations of the arguments $t$ and $t'$, and the new eigenfunction  problem can be posted:
\BY
&& \sqrt{\lambda_i(\overline{\Delta L})} \phi_i (t;\overline{\Delta L}) =\int\limits_{0}^{T_w}dt'\;G(t,t';\overline{\Delta
L})\phi_i(t';\overline{\Delta L}),
\EY
where $\phi_i (t;\overline{\Delta L})$ are eigenfunctions, and $\sqrt{\lambda_i(\overline{\Delta L})}$ are eigenvalues of the memory on the thermal
atomic ensemble. Note, these sets are different for each temperature.
\begin{figure}
\includegraphics[height=7.0cm]{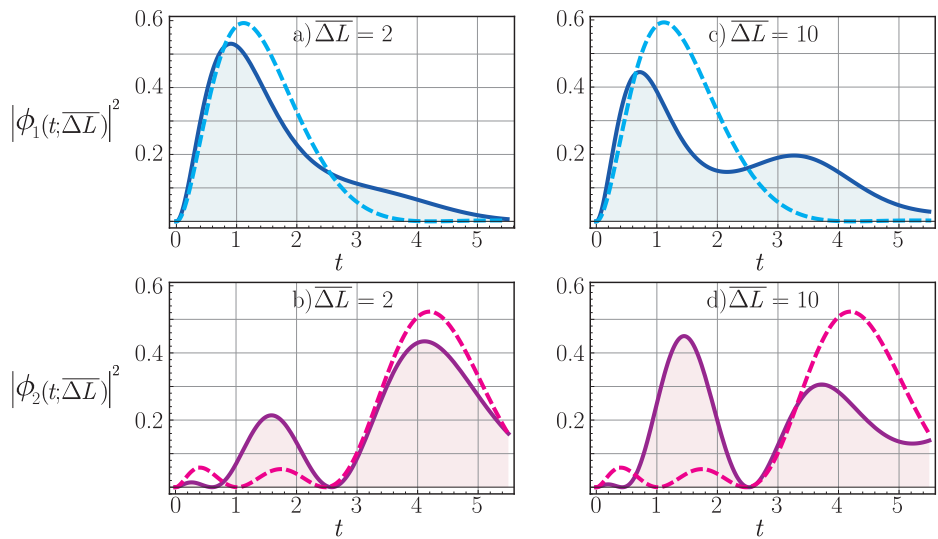}
\caption{The squares of eigenfunctions $\phi_1 (t;\overline{\Delta L})$ and $\phi_2 (t;\overline{\Delta L})$ of full memory cycle including storage
stage (solid curves). In frames a) and b) $\overline{\Delta L}=2$, in frames c) and d)  $\overline{\Delta L}=10$. Dashed curves are given for ease of
comparison and correspond to $\overline{\Delta L}=0$. Dimensionless time $t$ is given in units of $\Omega^{-1}$. } \L{Fig7}
\end{figure}

Fig. \ref{Fig7} shows the squares of the first (blue curve) and the second (purple curve) eigenfunctions for $\overline{\Delta L}=2$ (left column)
and $\overline{\Delta L}=10$ (right column). For comparison the analogous curves for motionless atoms are plotted by dashed lines. It can be seen
that the time profiles of intensities significantly changed. In particular, we take notice that profile of $|\phi_1(t; \overline{\Delta L})|^2$ at
$\overline{\Delta L}=10$ became alike the intensity profile of the output pulse for room-temperature atoms (see Fig. \ref{Fig6}b). However, the main
thing here is that for the new eigenfunctions the memory efficiency have changed considerably. Numerical calculation shows that after the
optimization the new efficiencies equal $\eta_1=94\%$ and $\eta_2=41\%$ at $\overline{\Delta L}=2$, and $\eta_1=74\%$ and $\eta_2=3\%$ at
$\overline{\Delta L}=10$. (Let us remind that before we got $\eta_1=87\%$, $\eta_2=46\%$ at $\overline{\Delta L}=2$, and $\eta_1=55\%$, $\eta_2=22\%$
at $\overline{\Delta L}=10$.) Although the value $\eta_2$ turns out to be lower than 50\% threshold of quantum memory, the proposed optimization
procedure enabled to increase $\eta_1$ significantly, even for large displacements $\overline{\Delta L}$.

\section{Conclusion}\L{V}

We have analyzed the impact of the longitudinal thermal motion on the operation of multimode quantum memory, and have shown that atomic mobility
yields the redistribution of spin excitations, which can be described in terms of the response functions. Redistribution of the excitations between
the response functions causes the distortion of the retrieved field profile: it turns out to be different from the incident one, and the disparity
increases with the temperature of the ensemble. As a result, the efficiency of the full memory cycle decreases.

It is important to note that even with significant displacements of atoms during the storage time, the memory remains to be quantum. Moreover, the
quantum nature of the memory is preserved even with the complete mixing of atoms in the cell. This once again confirms that the quantum state of
light is recorded on the collective of atoms but not on the individual ones, and the atomic movement in this collective, although it leads to
distortion of the mode structure of memory, but does not affect drastically the ability of the quantum storage.

Note, in some situations the longitudinal movement can cause not a decrease of the efficiency of the quantum memory, but the contrary, an increase of
it. For example, when one record a mode of the full memory cycle with a relatively small eigenvalue, thermal spin redistribution can lead to
population of readout modes with the larger eigenvalues that should result in the increase of the efficiency.

In this paper, we analyzed the role of the thermal motion in the model of high-speed quantum memory. The choice of the model is primarily caused by
the ability to ignore the displacements of the atoms on the writing and readout stages, setting them short. However, for other models of quantum
memory (adiabatic, QND, Raman), this statement of the problem is possible, if we assume that the atomic velocities are sufficiently small, whereas
the ratio between the storage time and interaction time is large. Different memory models provide different mode structures. For example, for the
adiabatic memory one can assume the minor disruption when the first mode is stored, because, as we have shown, the losses due to the thermal motion
depend on the shape of the response function which formed in the medium after writing.

We have proposed to optimize the memory taking into account the thermal motion at the storage, and fulfilled this optimization by finding the new
field profiles, which interact optimally with a mobile atomic ensemble.

One of the unexpected results is that the quantum memory at room temperature may be even more effective than with cold atoms. This is winning for the
practical application of the quantum memory on the atomic ensembles, in particular, it is essential for its scalability.

We can conclude that for the effective operation of quantum memory is important not only to preserve the spin excitations, but also to provide their
optimal spatial distribution.

\section{Acknowledgments}
The reported study was supported by RFBR (Grants No. 15-02-03656a and No. 13-02-00254a).

\appendix
\section{Eigenfunctions in the case of different durations of writing and retrieval}

According to Eq. (\ref{25}) the evolution of the signal field during its writing/readout is determined by the kernel of the integral equation
$G(t,t^\prime)$. As known, the initial set of the equations describes the system development only during the limited time interval, thereby we can
rewrite the kernel in the following way:
\BY
&&G(t,t^\prime)=G_0(t,t^\prime)\Theta_{{r}}(t)\Theta_{{w}}(t^\prime),
\EY
where the factors $\Theta_r(t)$ and $\Theta_w(t)$ are unequal to zero only during the interaction times:
\BY
&&\Theta_{r}(t)=1 \quad \mbox{at}\quad0<t<T_r\quad \mbox{and}\quad\Theta_{w}(t)=1 \quad \mbox{at}\quad 0<t<T_w.
\EY
Here $T_w$ and $T_r$ are durations of the writing and the retrieval, respectively. The function  $G_0(t,t^\prime)$ turn into the kernel
$G(t,t^\prime)$ at the stationary conditions, i.e. at $T_w,\;T_r\to\infty$. For the most models of quantum memory this function is symmetrical with
respect to permutation of its arguments, $G_0(t,t^\prime)=G_0(t^\prime,t)$. However, as one can see, for limited observation times the kernel
$G(t,t^\prime)$ preserve this property only at $T_w=T_r$. The presence of two different $\Theta$-functions destroys the symmetry of the kernel
relative to its arguments permutation $t\leftrightarrow t^\prime$. Nevertheless, it is possible to provide the symmetry with respect to $k
t\leftrightarrow t^\prime$, where $k=T_w /T_r$. Indeed, we can write $G(t,t^\prime)=\tilde G(kt,t^\prime)$, then we get
 \BY
&&\tilde G(k t,t^\prime)=\tilde G_0(kt,t^\prime)\Theta_{r}(k t)\Theta_{w}(t^\prime)= \tilde G_0(k t,t^\prime)\Theta_{w}(k
t)\Theta_{w}(t^\prime)=\tilde G(t^\prime,k t).
\EY
Therefore, the Schmidt decomposition for this kernel can be given as
\BY
&&G(t,t^\prime)=\tilde G(k t,t^\prime)=\sum_i\sqrt{k\;\lambda_i}\;\phi_i(k t)\phi_i(t^\prime),\L{15}
\EY
where $\lambda_i$ and $\phi_i(t)$ are the eigenvalues and the eigenfunctions of the kernel $\tilde G$, so that
\BY
&&\sqrt{k\lambda_i}\;\phi_i(k t)=\int_0^{T_w}dt^\prime\phi_i(t^\prime)\:\tilde G(k t,t^\prime).
\EY
The set of functions $\{\phi_i(t)\}_{i=1}^{\infty}$ is determined at the domain $[0,T_w]$ and satisfy the conditions:
\BY
&&\int_0^{T_w}dt\phi_i(t)\phi_j(t)=\delta_{ij},\qquad\sum_i\phi_i(t)\phi_i(t^\prime)=\delta(t-t^\prime).\L{16}
\EY
The functions $\varphi_i(t)=\sqrt k \;\phi_i(k t)$ are given at the domain $[0,T_r]$, and, one can easily see, also meet the conditions of
orthonormality and completeness:
\BY
&&\int_0^{T_r}dt\varphi_i(t)\varphi_j(t)=\delta_{ij},\qquad\sum_i\varphi_i(t)\varphi_i(t^\prime)=\delta(t-t^\prime).\L{17}
\EY

The sets of functions $\{\phi_i(t)\}_{i=1}^{\infty}$ and $\{\varphi_i(t)\}_{i=1}^{\infty}$ can be interpreted as eigenfunctions at the writing and
the retrieval stages, correspondingly, and it is obvious, they can be obtained from each other by a simple scaling the time with the coefficient $k$.


\begin{thebibliography}{100}
%
\bibitem{HammererReview} K. Hammerer, A.S. Sorensen, and E.S. Polzik, Rev. Mod. Phys. \textbf{82}, 1041 (2010).
%
\bibitem{LvovskyREview}  A. I. Lvovsky, B. C. Sanders, and W. Tittel, Nat.Photonics \textbf{3}, 706 (2009).
%
\bibitem{Simon2007} J. Simon, H. Tanji, J. K. Thompson, and V. Vuleti\'{c}, Phys. Rev. Lett. \textbf{98}, 183601 (2007).
%
\bibitem{Chou2007} C.-W. Chou, L. Laurat, H. Deng, K. S. Choi, H. de Riedmatten,D. Felinto, and H. J. Kimble, Science \textbf{316}, 1316 (2007).
%
\bibitem{Chen2008} Y.-A. Chen, S. Chen, Z. S. Yuan, B. Zhao, C. S. Chuu, J. Schmiedmayer, and J.-W. Pan, Nature Physics \textbf{4}, 103 (2008).
%
\bibitem{Tittel2010} W. Tittel, M. Afzelius, T. Chaneliere, R.L. Cone, S. Kroll, S.A. Moiseev, M. Sellars, Laser Photon. Rev. \textbf{4}, 244 (2010).
%
\bibitem{Buchler2011} M. Hosseini, G. Campbell, B. M. Sparkes, P. K. Lam, and B. C. Buchler, Nature Physics \textbf{7}, 794 (2011).
%
%
\bibitem{Polzik2011} K. Jensen, W. Wasilewski, H. Krauter, T. Fernholz, B.M. Nielsen, J.M. Petersen, J.J. Renema,
M.V. Balabas, M.Owari, M.B. Plenio, A. Serafini, M.M. Wolf, C.A. Muchik, J.I. Cirac, J.H. M\"{u}ller and E.S Polzik, J. Phys.: Conf. Ser.
\textbf{66}, 275 (2012).
%
\bibitem{Camacho2008} R.M. Camacho, P.K. Vudyasetu and J.C. Howell, Nature Photonics \textbf{290}, 103-106 (2008).
%
\bibitem{Novikova2011} N.B. Phillips, A.V. Gorshkov, and I. Novikova,  Phys. Rev. A \textbf{83}, 063823 (2011).
%
\bibitem{Tabosa2014} A.J.F. de Almeida, J. Sales, M.-A. Maynard, T. Laupr\'{e}tre, F. Bretenaker, D. Felinto, F. Goldfarb, and J. W. R. Tabosa,
Phys. Rev. A \textbf{90}, 043803 (2014).
%
\bibitem{Lvovsky2014} T. Brannan, Z. Qin, A. MacRae, and A. I. Lvovsky,  Optics Letters \textbf{39}, 18 (2014).
%
%
%
%
%
%
%
%
%
%
\bibitem{Novikova2012} I. Novikova, R.L. Walsworth, and Y. Xiao, Laser Photonics Rev. \textbf{6}, 3, 333-353 (2012).
%
\bibitem{Tittel2014} N. Sinclair, E. Saglamyurek, H. Mallahzadeh, J. A. Slater, M. George, R. Ricken,
M. P. Hedges, D. Oblak, C. Simon, W. Sohler, W. Tittel, Phys. Rev. Lett. \textbf{113}, 053603 (2014).
%
\bibitem{Polzik2015} J. Borregaard, M. Zugenmaier, J. M. Petersen, H. Shen, G. Vasilakis, K. Jensen, E. S. Polzik, A. S. Sorensen,
Room temperature quantum memory and scalable single photon source based on motional averaging, arXiv:1501.03916 [quant-ph] (2015).
%
\bibitem{Gorshkov1} A.V. Gorshkov, A. Andr\'{e}, M.D. Lukin, and A.S. Sorensen, Phys. Rev. A, \textbf{76}, 033804 (2007).
%
\bibitem{Gorshkov2} A.V. Gorshkov, A. Andr\'{e}, M.D. Lukin, and A.S. Sorensen,  Phys. Rev. A, \textbf{76}, 033805 (2007).
%
\bibitem{Vasilyev} D.V. Vasilyev, I.V. Sokolov, and E.S. Polzik, Phys. Rev. A, \textbf{77}, 020302 (2008).
%
\bibitem{GrodeckaGrad} E. Zeuthen, A. Grodecka-Grad, and A.S. Sorensen, Phys. Rev. A, \textbf{84}, 043838 (2011).
%
\bibitem{Nunn2008} J. Nunn, K. Reim, K.C. Lee et al., Phys. Rev. Lett. \textbf{101}(26), 260502 (2008).
%
\bibitem{Kimble} K. S. Choi, H. Deng, J. Laurat, and H. J. Kimble, Nature \textbf{452}, 67 (2008).
%
\bibitem{Kuzmich} R. Zhao, Y. O. Dudin, S. D. Jenkins, C. J. Campbell, D. N. Matsukevich, T. A. B. Kennedy, and A.
Kuzmich, Nature Phys. \textbf{5}, 100 (2009).
%
\bibitem{Schmiedmayer} B. Zhao, Y. A. Chen, X.-H. Bao, T. Strassel, C. S. Chuu, X.-M. Jin,
J. Schmiedmayer, Z. S. Yuan, S. Chen, and J.W. Pan, Nature Phys.\textbf{5}, 95 (2009).
%
\bibitem{Vuletic} H. Tanji, S. Ghosh, J. Simon, B. Bloom, and V. Vuleti\'{c}, Phys. Rev. Lett. \textbf{103}, 043601
(2009).
%
%
%
\bibitem{LandauLifshitz} L.D. Landau, E.M. Lifshitz, Statistical Physics, Part 1. Vol. 5, Butterworth–Heinemann (1980)
%
%
%
\bibitem{LauratQM1} L. Veissier, A. Nicolas, L. Giner, D. Maxein, A.S. Sheremet, E. Giacobino, J. Laurat, Optics Letters \textbf{38}, 5, 712 (2013).
%
\bibitem{LauratQM2} A. Nicolas, L. Veissier, L. Giner, E. Giacobino, D. Maxein, J. Laurat,  Nature Photonics \textbf{8}, 234–238 (2014).
%
%
\bibitem{HSQM1} T.Y. Golubeva, Y.M. Golubev, O.Mishina, A. Bramati, J. Laurat, E.Giacobino, Phys.Rev. A, \textbf{83}(5), 053810 (2011).
%
\bibitem{HSQM2} T.Y. Golubeva, Y.M. Golubev, O. Mishina, A. Bramati, J. Laurat, E. Giacobino, Eur. Phys. J. D, \textbf{66}, 275 (2012).
%
\bibitem{HS2014} K. Tikhonov, K. Samburskaya, T.Y. Golubeva, Y.M. Golubev, Phys. Rev. A., \textbf{89}, 013811 (2014).
%
%
\bibitem{Golubev2015} T. Golubeva, Yu. Golubev, Role of the efficiency in multimode quantum memory, to be published in 2015.
%
\bibitem{Balabas1} M.V.Balabas, T.Karaulanov, M.P.Ledbetter, D.Budker, Phys. Rev. Lett. \textbf{105}, 070801 (2010).
%
\bibitem{Balabas2}M.V. Balabas ,K. Jensen, W. Wasilewski, H. Krauter, L.S. Madsen, J.H. Muller, T. Fernholz, E.S. Polzik,
Optics Express \textbf{18}, 6, 5825-5830 (2009).
%
\bibitem{Balabas3} M. T. Graf, D. F. Kimball, S. M. Rochester, K. Kerner, C. Wong, and D. Budker, E. B. Alexandrov and M. V. Balabas,
Phys. Rev. A \textbf{72}, 023401 (2005).
%
\bibitem{Nunn2007} J. Nunn, I.A. Walmsley, M.G. Raymer, K. Surmacz, F.C. Waldermann, Z. Wang, D. Jaksch, Phys. Rev. A \textbf{75}, 011401 (2007).
%

\end{thebibliography}
\end{document}